\documentclass[]{aastex631}

\begin{document}

\title{The GMRT High Resolution Southern Sky Survey for pulsars and transients -- VI: Discovery of nulling, localisation and timing of PSR J1244$-$4708}

\author{S.~Singh}
\affiliation{National Centre for Radio Astrophysics, Tata
Institute of Fundamental Research, Pune 411 007, India}

\author{J.~Roy}
\affiliation{National Centre for Radio Astrophysics, Tata
Institute of Fundamental Research, Pune 411 007, India}

\author{Shyam~S.~Sharma}
\affiliation{National Centre for Radio Astrophysics, Tata
Institute of Fundamental Research, Pune 411 007, India}

\author{B.~Bhattacharyya}
\affiliation{National Centre for Radio Astrophysics, Tata
Institute of Fundamental Research, Pune 411 007, India}

\author{S.~Kudale}
\affiliation{National Centre for Radio Astrophysics, Tata
Institute of Fundamental Research, Pune 411 007, India}

%% Note that the \and command from previous versions of AASTeX is now
%% depreciated in this version as it is no longer necessary. AASTeX 
%% automatically takes care of all commas and "and"s between authors names.

%% AASTeX 6.31 has the new \collaboration and \nocollaboration commands to
%% provide the collaboration status of a group of authors. These commands 
%% can be used either before or after the list of corresponding authors. The
%% argument for \collaboration is the collaboration identifier. Authors are
%% encouraged to surround collaboration identifiers with ()s. The 
%% \nocollaboration command takes no argument and exists to indicate that
%% the nearby authors are not part of surrounding collaborations.

%% Mark off the abstract in the ``abstract'' environment. 
\begin{abstract}
Many pulsars in the known population exhibit nulling, which is characterised by a sudden cessation and subsequent restoration of radio emission. In this work, we present the localization, timing, and emission properties of a GHRSS discovered pulsar J1244-4708. Moreover, we find that this pulsar shows nulling with a nulling fraction close to $60\%$. A quasi-periodicity is also seen in the nulling from this pulsar with two timescales. We demonstrate the broadband nature of nulling in this pulsar using simultaneous observations in band-3 (300-500 MHz) and band-4 (550-750 MHz) with the uGMRT. We also present a comparison of the efficiency of various search approaches such as single pulse search, Fast Folding Algorithm (FFA) based search, and Fast Fourier Transform (FFT) based search to search for nulling pulsars. We demonstrated that the FFA search is advantageous for detecting extreme nulling pulsars, which is also confirmed with multiple epochs of observations for the nulling pulsars using the GMRT.

\iffalse{One of the objectives of the pulsar surveys is to find new pulsars with interesting emission properties. Rotating Radio Transients (RRATs) and nulling pulsars are objects that emit a periodic yet erratic signal. The radio emission from these objects frequently ceases and then resumes. This process of stopping and restarting of the radio emission is expected to be connected with the radio emission mechanism. In this paper, we compare the efficiency of various search approaches for nulling pulsars and RRATs (single pulse search, Fast Folding Algorithm (FFA) based search, and Fast Fourier Transform (FFT) based search). We also present the localization, timing, and emission properties of a GHRSS pulsar J1244-4708. This pulsar shows nulling with a nulling fraction close to 60$\%$. This pulsar also shows periodic nulling with two-time scales of periodicity. We demonstrate the broadband nature of nulling in this pulsar using a simultaneous observation in band-3 (300-500 MHz) and band-4 (550-750 MHz) of the uGMRT.}\fi

\end{abstract}

%% Keywords should appear after the \end{abstract} command. 
%% The AAS Journals now uses Unified Astronomy Thesaurus concepts:
%% https://astrothesaurus.org
%% You will be asked to selected these concepts during the submission process
%% but this old "keyword" functionality is maintained in case authors want
%% to include these concepts in their preprints.

%% From the front matter, we move on to the body of the paper.
%% Sections are demarcated by \section and \subsection, respectively.
%% Observe the use of the LaTeX \label
%% command after the \subsection to give a symbolic KEY to the
%% subsection for cross-referencing in a \ref command.
%% You can use LaTeX's \ref and \label commands to keep track of
%% cross-references to sections, equations, tables, and figures.
%% That way, if you change the order of any elements, LaTeX will
%% automatically renumber them.
%%
%% We recommend that authors also use the natbib \citep
%% and \citet commands to identify citations.  The citations are
%% tied to the reference list via symbolic KEYs. The KEY corresponds
%% to the KEY in the \bibitem in the reference list below. 

\section{Introduction} \label{sec:intro}
Since the discovery of the first pulsar \citep{Hewish}, there has been a consistent effort to understand the radio emission physics from these objects. Even after decades of rigorous work on both the theoretical and observational front, there are still many questions regarding radio emission physics that are not completely answered (\citet{Beskin_2018}, \citet{Mitra_2017}, and \citet{Cerutti_2016}). The stable profile of pulsars, obtained after folding thousands of single pulses, contains information about the average beam shape and geometry of the pulsar, whereas the single pulse properties like microstructures \citep{micro_craft}, subpulse drifting \citep{drifting_drake}, mode changing \citep{modes_bartel}, nulling \citep{Nulling_backer}, and intermittency \citep{Konar_2019} are the features that are used to understand the exact physics of pulsar radio emission. \citet{Nulling_backer} was the first to report the nulling phenomena in pulsars, which is a cessation of normal pulsed emission for a period of time. Till now, over 200\footnote{http://www.ncra.tifr.res.in/~sushan/null/null.html} pulsars are known to show nulling. \citet{Konar_2019} indicated the lack of  dedicated search for nulling in the current population and a careful investigation and search for nulling in known pulsar population are required to truly estimate the number of nulling pulsars.  Moreover, a major fraction of the current pulsar population has not been studied for their single pulse properties due to the sensitivity limitations of radio telescopes.\\

There are many pulsar surveys trying to increase the pulsar population by using more sensitive telescopes and new search techniques (e.g. \citet{GPPS}, \citet{SMART}, \citet{SUPERB}, \citet{GHRSS1}, \citet{GHRSSIV}, \citet{TRAPUM}). Along with normal and millisecond pulsars, many new exciting time-domain transients have been discovered by these surveys, including ultra-long period pulsars (e.g. \citet{76s_psr}), Rotating radio transients (RRATs, e.g. \citet{Maura_RRAT} and \citet{RRAT_discovery}), and many new Fast Radio Bursts (FRBs, e.g. \citet{SUPERB_FRB} and \citet{FRB_review}). One of the main objectives of these surveys is to discover more pulsars with interesting emission features that can help us to constrain the radio emission mechanism. Discovering new nulling pulsars and RRATs poses a significant challenge. The population of RRATs has been defined as sources of repeating signals that can be more efficiently detected in single pulses than the periodicity searches, though they have inherent periodicity \citep{konar_RRAT}. Some nulling pulsars show very long nulls lasting for a few tens of minutes, which is larger than the integration time per pointing in many major pulsar surveys. There is also a question of a suitable search method for nulling pulsars, especially for pulsars with severe nulling. Sufficient bright pulsars can be found in a single pulse search, but a periodicity search will be needed for a fainter population of pulsars with extreme nulling. Fast Fourier Transform (FFT) based periodicity searches (e.g. \texttt{PRESTO}, \citet{Ransom_2002}) require a signal to be consistent in its periodicity. While the Fast Folding Algorithm (FFA, \citet{staelin}) search method searches for periodic signals by folding the time series at trial periods and can capture any signal that appears in the folded profile. A detailed comparison of these two periodicity search methods is required to select the best-suited methodology to search for pulsars with long nulls.\\ 

The nulling phenomenon can be characterized by three parameters: nulling fraction, null as well as burst lengths, and periodicity of nulling.  The nulling fraction measures the fraction of observation time when there is no signal detected from the pulsar. The null and burst lengths provide a measure of the typical timescales of the nulling and continuous pulsed emission respectively. Sometimes nulling appears to be quasi-periodic (e.g. \citet{nulling_basu} and \citet{periodic_rankin}) and an approximate period of the nulling can be measured. These inputs from nulling phenomena are crucial to evaluate the radio emission models that attempt to explain nulling in pulsars. There are two popular classes of models that explain the nulling in pulsars on the basis of the radio emission mechanism. One class is based on the pulsar death Valley \citep{Nulling_age}. Pulsars stop emitting in radio when their magnetic field strength is not enough to sustain pair production. This class of models proposes that nulling pulsars have magnetic fields close to this boundary value and these pulsars only emit when conditions are favorable otherwise they remain in a null state \citep{nulling_death_line}. Though this model looks interesting and feasible, many nulling pulsars are located far away from the pulsar death line, thus having enough field strength to sustain pair production \citep{Konar_2019}. The second class of models relies on changes in the emission state \citep{mode_change_eg} or the loss of coherence in the radio emission process \citep{coherence_loss_eg}. One of the proposed reasons behind this loss of coherency is the change in the surface magnetic field configuration \citep{nulling_minhajur} due to surface current on the polar cap \citep{state_change_machs}. This change in field configuration can also explain the mode change (a phenomenon where the mean pulse profile abruptly switches between two or more quasi-stable states) associated with the nulling. The quasi-periodic nulling seen in many pulsars and the consistent lengths of nulls and bursts can also be explained by this model \citep{nulling_minhajur}. \\

We are performing a low-frequency pulsar survey with the GMRT radio interferometer, called GMRT High Resolution Southern Sky (GHRSS\footnote{http://www.ncra.tifr.res.in/~bhaswati/GHRSS.html}) survey. The survey, currently operating in phase-II over a frequency range of 300$-$500 MHz and targeting GMRT sky within $-$20 degree $<$ declination $<$ $-$54 degree, has found 28 new pulsars till now (\citet{GHRSS1}, \citet{GHRSS2}, \citet{GHRSS3}, \citet{GHRSSIV}, and \citet{localization}) including two nulling pulsars. In this paper, we present localization, timing, and discovery of nulling for PSR J1244-4708, a pulsar discovered by the GHRSS survey. We discuss the efficiency of single pulse search and periodicity search methods to find nulling pulsars in section \ref{sec:detect}. Localization and timing of the GHRSS pulsar J1244$-$4708 is presented in section \ref{sec:loc}. We discuss the folded profile of J1244$-$4708 and its frequency evolution in section \ref{sec:profile}. We present the various nulling properties of this pulsar in section \ref{sec:nulling} and summarize this work in section \ref{sec:sum}.

\section{Detectability of nulling pulsars in periodicity search}\label{sec:detect}
Periodicity search methods are used to detect a periodic signal present over the observing duration consisting of fainter single pulses. The most sensitive periodicity search for non-accelerated signals is the Fast Folding Algorithm (FFA) search \citep{RIPTIDE}. In this method, the detection is done on the folded profile. Use of a periodicity search will be only useful if the folded profile S/N (signal-to-noise ratio) is larger than the single pulse S/N. If the peak S/N of a single pulse is S, then the peak S/N of the folded profile after folding N such pulses will be given by,
$$S_{fold} = S \sqrt{N}$$
Now, if we consider a nulling fraction of $f_{null}$ in the time series (i.e. only $(1-f_{null})N$ pulses are having signal), then the S/N of the profile will become,
\begin{equation}
 S_{fold}=(1-f_{null})S\sqrt{N}   
\end{equation}

The signal will be detected in folded profile if $S_{fold} > S_{threshold}$, where $S_{threshold}$ is the threshold for detection,
$$S(1-f_{null})\sqrt{N} > S_{threshold}$$
The number of pulses N in the observation can be written as $\frac{\tau}{P}$, where $\tau$ is observation duration and $P$ is the period of the pulsar,
$$(1-f_{null})\sqrt{\frac{\tau}{P}}>\beta$$
Here $\beta$ is the ratio of the detection threshold and the average single pulse S/N.
\begin{equation}
 \tau>\frac{{\beta}^2}{(1-f_{null})^2}P   
\end{equation}

This relation tells that all nulling pulsars irrespective of the severity of their nulling can be discovered in periodicity search, given that we have a sufficiently long observation. One can also calculate the maximum nulling fraction of a pulsar for a particular period and time of observation, which can be detected better in a folded profile than in a single pulse search. Considering the typical pointing duration of around 10 minutes for the GHRSS survey and a pulsar period of 1 second, the pulsar will be better detected in folded profile if the nulling fraction is less than $96\%$.\\

In the phenomena of nulling, the signal appears and disappears from pulse to pulse. Though the signal is periodic, the absence of a pulse weakens the overall signal strength and also affects the performance of search methods that are strictly based on periodicity. The FFT-based search methods are based on the periodicity of the signal and the detectability in the power spectra depends on the consistency/regularity of the periodic signal. In FFA search, the time series is folded at all possible trial periods and detection is essentially done on the folded profiles. The features that appears in the folded profile with underlying periodicity (but having lack of regular emission) can be detected by the FFA search. This fundamental difference between the two periodicity search methods warrants a detailed comparison of their performance on nulling signals. It should be noted that the superiority of FFA search over the FFT search for all non-accelerated signals is already established by \citet{RIPTIDE} and \citet{GHRSS3}. In this work we are looking further for any change in the relative performance of these two search methods caused by the nulling of the pulsed emission.\\

To simulate nulling pulsars, we injected pulses in the white noise time series with a length of 10 minutes (similar to the GHRSS observing pointing duration) at a period of 2 s. We also simulated a telescope-like noise time series with rednoise conditions similar to the GHRSS survey \citep{GHRSS3}. We used this noise time series to account for the rednoise condition in the GMRT time-domain data allowing a more realistic comparison. We injected a given pulse shape at a given rotation period in a time series. We used a period of 2 s and a Gaussian pulse shape having a full-width half maxima (FWHM) equivalent to 1$\%$ of the rotation period. To simulate the nulling pulsar with a given nulling fraction, we used a random number following a uniform distribution in a range [0.0,1.0]. We generated this number for each pulse cycle and if the value is larger than the nulling fraction, then we injected a pulse in that period cycle otherwise leave that portion of the noise time series unaltered. We also used another random number following a Gaussian distribution to introduce variation in the strengths of injected pulses. We varied the nulling fraction of the signal in the range zero to one.\\

\begin{figure}
    \centering
    \includegraphics[height=9 cm, width=13 cm]{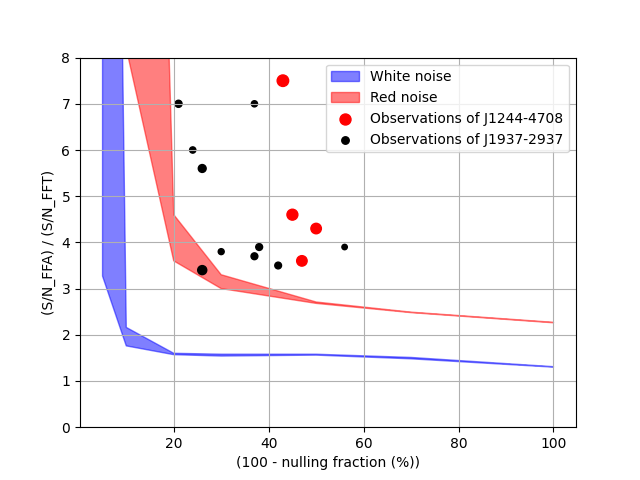}

    \caption{Relative performance of FFA and FFT search as a function of nulling fractions for the pulsars considering white noise and red noise. For pulsars with high nulling fractions, the FFA search is advantageous. This is also supported by multiple epochs of observations for the GHRSS pulsars J1244-4708 and J1937-2937. The marker size is proportional to the observation duration.}
    \label{fig:fig1}
\end{figure}

Fig. \ref{fig:fig1} shows the relative performance of the FFA and FFT searches as a function of the nulling fraction. Though, in general the relative performance of the FFA search is better for all nulling fractions, a boost in the relative performance of the FFA search is seen in the white noise cases when the nulling fraction is very high (larger than 90\%). In the presence of rednoise, we see the boost in relative performance at lower nulling fractions (as low as 50\%) as well. We also use the nulling pulsar J1936$-$30 (now J1937$-$2937) reported in \citet{GHRSSIV} and another nulling pulsar J1244$-$4708, whose properties are first time getting reported in this work, to compare the two search methods. These data points from real nulling pulsars are loosely following the trend of higher FFA S/N for larger nulling fractions. The ratio of FFA S/N and FFT S/N depends on the rednoise conditions in that particular observation. Most of the data points are above the trend in the presence of simulated rednoise, indicating that the rednoise conditions in those observations are more severe than the rednoise parameters we used for the simulation\\  

To summarize, FFA search is always better for any non-accelerated signal but an additional advantage is seen for pulsars with extreme nulling. This advantage gets amplified in the presence of rednoise. So, the extreme nulling pulsars give us one more reason to use FFA search to find these isolated systems.

\section{Localization and timing of PSR J1244$-$4708} \label{sec:loc}
\begin{figure}
    \centering
    \includegraphics[height=8 cm, width=12 cm]{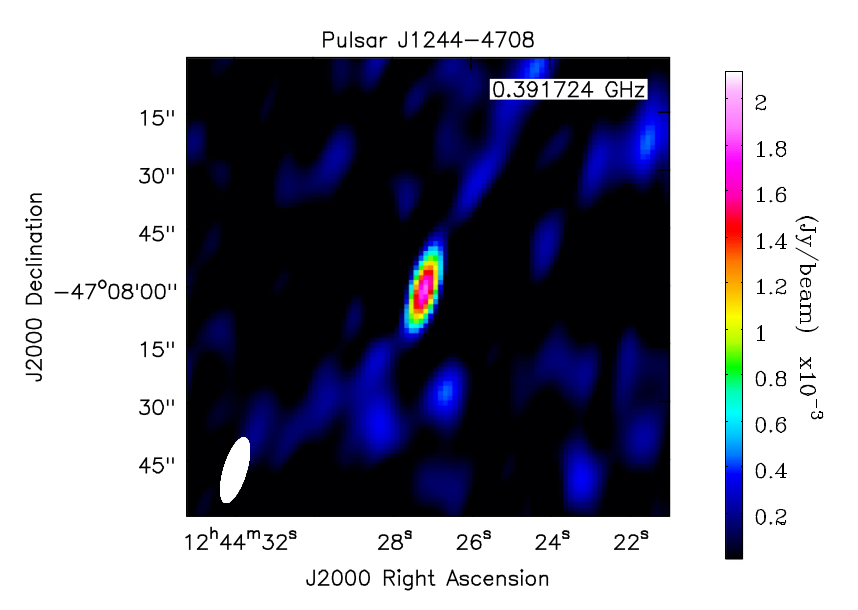}
    \caption{An image with the uGMRT data showing the location of PSR J1244$-$4708. This is a 1.3 mJy source detected at 10 $\sigma$ in the wide-band image at an offset of $20'$ from the GHRSS pointing center.}
    \label{fig:fig2}
\end{figure}

The GHRSS survey in phase-II with the upgraded GMRT (uGMRT) discovered PSR J1244$-$4708 in the FFT search (Bhattacharyya et al. 2023 in prep.). This is a long-period pulsar with a period of 1.4114 s and a Dispersion Measure (DM) of 75.0 $pc~cm^{-3}$. The GHRSS survey uses the incoherent array (IA) mode of observations, in order to increase the sky coverage per pointing. In phase-II, the IA beam of GMRT has a Half Power Beam Width (HPBW) of 64$'$ at 400 MHz. Hence each in-beam discovery can have an uncertainty of $\sim\pm32'$ in their location. Accurate localization is needed to follow up a newly discovered pulsar with the narrow and more sensitive phased array (PA) beam of the GMRT. We used the multiple phased array beamformation method described by \citet{localization} to localize J1244$-$4708. We imaged the field with the wide-band uGMRT data in which the pulsar was discovered to extract the point sources and their locations. Then, we formed PA beams using the legacy GMRT baseband data having 33 MHz bandwidth at each of these locations and check for pulsed emission in each beam. One of the point sources in this field is likely to be pulsar and the PA beam formed at that location is expected to give $\sqrt{N_{ant}}$ times more S/N than the IA beam detection, where $N_{ant}$ is the number of antennas used in the beam formation. We detected the pulsar with a very good S/N ($\sim$ 4 times of the IA beam) at one of the point sources in the field of view (FOV), located at a $\sim 20'$ offset from the GHRSS discovery pointing centre. Fig. \ref{fig:fig2} shows this point source with 1.3 mJy flux, detected at $10\sigma$ significance. \iffalse{The detection significance of the simultaneous PA and IA beam for this pulsar are shown in Fig. \ref{fig:fig4}. The PA beam in this image exhibits much cleaner detection of the pulsar in comparison with the IA beam detection.}\fi We get very clear detection of the pulsar with clear nulls at this location, confirming the pulsating nature of this point source.\\

This same field also contains a GHRSS millisecond pulsar J1242$-$4712 \citep{localization}. Due to the presence of this millisecond pulsar, this field was routinely observed for follow-up timing enabling us to use these data (i.e. IA beam) for the timing of PSR J1244$-$4708. We folded the data with \texttt{prepfold} and then used \texttt{get\_TOA.py} of \texttt{PRESTO} \citep{Ransom_2002} to calculate TOAs (time of arrivals) and then we used the pulsar timing software \texttt{TEMPO2} \citep{TEMPO2} to derive the timing model. We used the accurate location (12h44m27.2(5)s, -47d08'00(12)") obtained from the localization in the pulsar ephemeris. We fitted for position, period, and period derivative and were able to obtain phase connected timing solution for this pulsar. The timing model for this pulsar for an observation span of 577 days is given in table \ref{table:1}. The post-fit timing residuals are shown in Fig. \ref{fig:fig4}. The derived parameters like the estimate of surface magnetic field strength ($B_{s}$ = $1.2 \times 10^{12}$ G), the characteristic age ($\tau$ = 22.3 MYr), and the spin-down energy ($\dot{E}=1.4\times 10^{31}$ $erg~s^{-1}$) are also listed in the table. After noticing the signature of nulling in the PA beam, we followed up this pulsar with PA beam using GMRT wideband receivers in band-3 (300$-$400 MHz) and band-4 (550$-$750 MHz) of uGMRT, with an aim of detailed study of emission features of this pulsar. The results from these studies are reported in the subsequent sections.

\iffalse{\begin{figure}
    \centering
    \includegraphics[height=9 cm, width=16 cm]{IA_PA_detection.png}
    \caption{Detection of the pulsar J1244-4708 in simultaneous Incoherent Array (IA) beam and the Phased Array (PA) beam pointed at the location shown in Figure \ref{fig:fig2}. The PA beam has significantly strong detection with clear signature of mulling in comparison to the IA beam.}
    \label{fig:fig3}
\end{figure}}\fi

\begin{figure}
    \centering
    \includegraphics[height=9 cm, width=14 cm]{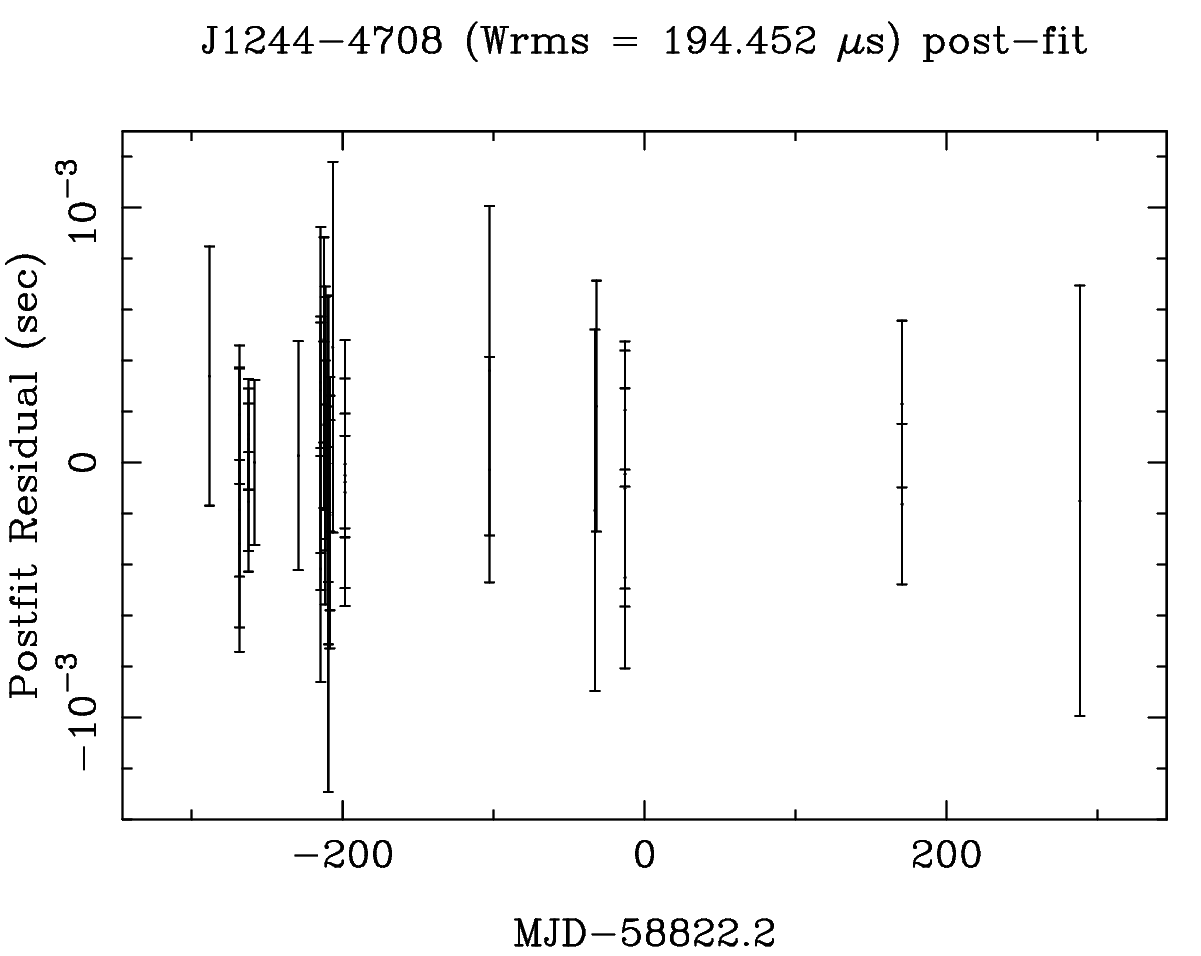}
    \caption{Timing residual for J1244-4708 at 400 MHz with the uGMRT observations. The data points span a range of 577 days.}
    \label{fig:fig4}
\end{figure}

\begin{deluxetable*}{cc}
\tablenum{1}
\tablecaption{Timing parameters of the pulsar J1244$-$4708}
\tablewidth{0pt}
\tablehead{
\colhead{Parameter} & \colhead{Value}}

\startdata
Pulsar Name & J1244$-$4708 \\
Right Ascension (J2000, h:m:s) & 12:44:27.236(3)\\
Declination (J2000, d:m:s) & -47:08:01.9(1) \\
Rotational frequency, F0 ($s^{-1}$)& 0.7084996307(1)\\
Frequency Derivative, F1 ($s^{-1}s$) & $-5.007(7)\times10^{-16}$\\
DM ($pc-cm^{-3}$) & 75.0174(3)\\
Period Epoch (MJD) & 58559.79799 \\
DM epoch (MJD) & 58522\\
Timing span (MJD) & 58533-59110\\
Number of TOAs & 43 \\
RMS timing residual ($\mu s$) & 194.45 \\
Solar system ephemeris model & DE405 \\
UNITS      &    TCB \\
Clock correction procedure & TT(TAI) \\
\hline
Derived parameters\\
\hline
Characteristic age & 22.3 MYr \\
Surface Magnetic field ($B_s$) & 1.2$\times 10^{12}$ G\\
Spin down energy ($\dot{E_{rot}}$) & 1.4$\times 10^{31}~erg~s^{-1}$\\
\enddata
\end{deluxetable*}
\label{table:1}

\section{Profile evolution of J1244$-$4708} \label{sec:profile}

\begin{figure}
   \centering
    \includegraphics[height=8 cm, width=18 cm]{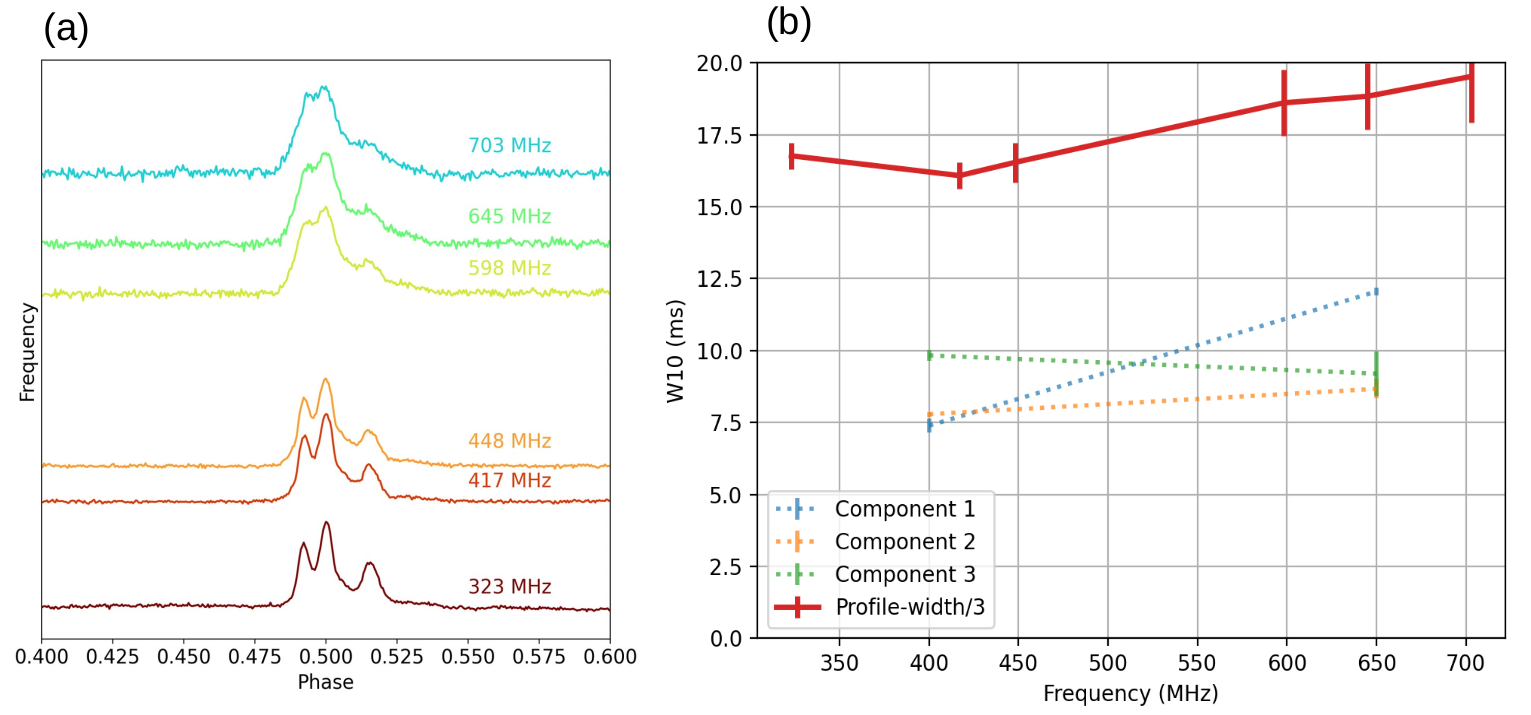}
    \caption{Frequency evolution of the average pulse profile for PSR J1244-4708. Panel (a) shows the pulse profile at multiple subbands of band-3 (300-500 MHz) and band-4 (550-750 MHz) uGMRT observations. All the profiles are individually normalized to the same height. The three clearly visible components at lower frequencies are getting merged at higher frequencies. Panel (b) shows the evolution of the widths of three components and the full profile as a function of frequency. We have plotted the W10/3 of the full profile to accommodate it with the measurements of individual component widths. Since the component fitting was done on the profiles obtained from the full bandwidth of band-3 and band-4, we have single measurements of component widths for each of the two bands.}
    \label{fig:fig5}
\end{figure}

 We used band-3 (300$-$500 MHz) and band-4 (550$-$750 MHz) phased array (PA) observations of this pulsar to study the frequency evolution of its profile. This pulsar shows three clear components in the folded profile and one faint component at the trailing end of the profile. We divided the 200 MHz bandwidth of both band-3 and band-4 data into three subbands to quantify the frequency evolution of the profile. Panel (a) of Fig.  \ref{fig:fig5} shows the profile at the central frequencies of different subbands. While there are three distinct components at lower frequencies, these components start merging with increasing frequencies. The merging of these components hints that the widths of the components are increasing with increasing frequencies. The Gaussian fitting of components revealed one more component between the second and third components. So, a total of five Gaussian components can fit in the profile. We fitted five Gaussian components in the profiles obtained from the full bandwidths of band-3 and band-4 observations, in order to get good S/N in the folded profile. We find that the locations of these components and hence separation between them remain unchanged (within the binning accuracy) across the frequency range. We measured the W10 (width at $10\%$ height of the peak) of the fitted Gaussian shapes associated with the three prominent components in the profile for both bands. We see that these widths are either increasing with increasing frequency or remain unchanged within the error bars. We plot the evolution of the W10 of the full profile along with the frequency evolution of the widths of three prominent components in panel (b) of Fig. \ref{fig:fig5}. The radius-to-frequency mapping relation states that higher frequencies are generated at lower heights and hence profile and its components should have smaller widths at higher frequencies \citep{RFM_book}. The evolution of width of both the full profile and the individual components of the pulsar J1244$-$4708 is opposite to what is expected from the radius-to-frequency relation, though the frequency coverage of our observation (300$-$750 MHz) is not enough to robustly conclude this. In fact, a significant fraction of the pulsar population deviates from the trend of smaller profile widths at higher frequency \citep{RFM}.\\

\section{Nulling properties of J1244$-$4708} \label{sec:nulling}
\begin{figure}
   \centering
    \includegraphics[height=8 cm, width=18 cm]{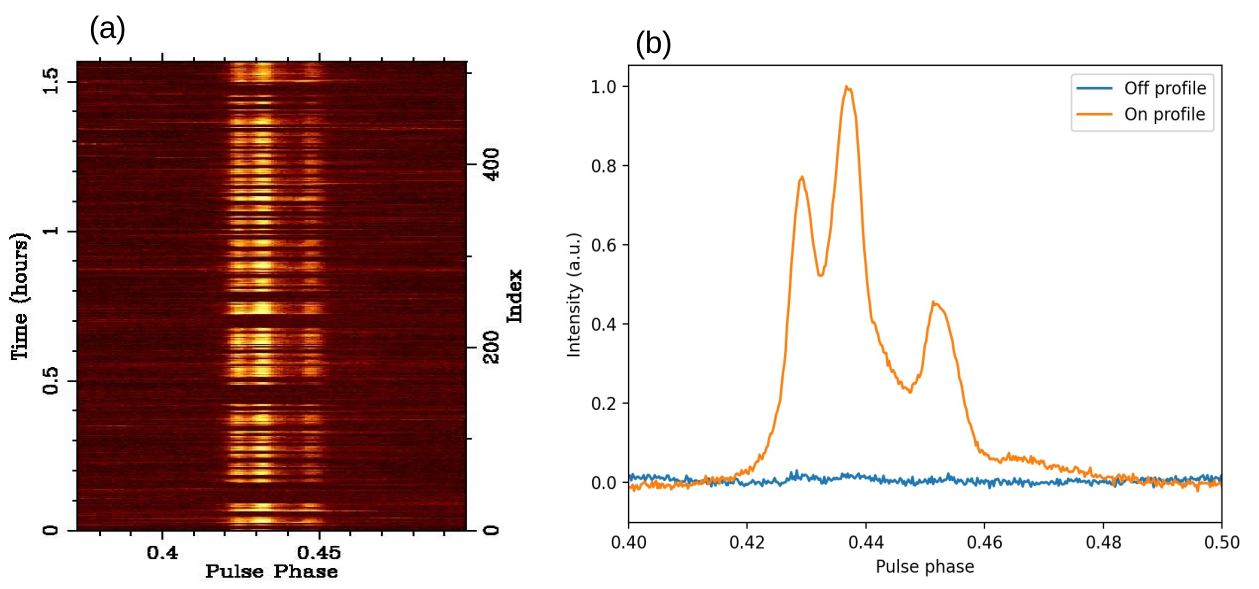}
    \caption{Nulling in band-3 (300$-$500 MHz) observation of J1244-4708 with the uGMRT. Panel (a) shows the pulse sequence and panel (b) shows the on and off profile of this pulsar. The off profile obtained from the nulls does not show a significant signal from the pulsar. This indicates that pulses that are nulling do not have any low-level emission.}
    \label{fig:fig6}
\end{figure}

\begin{figure}
   \centering
    \includegraphics[height=8 cm, width=18 cm]{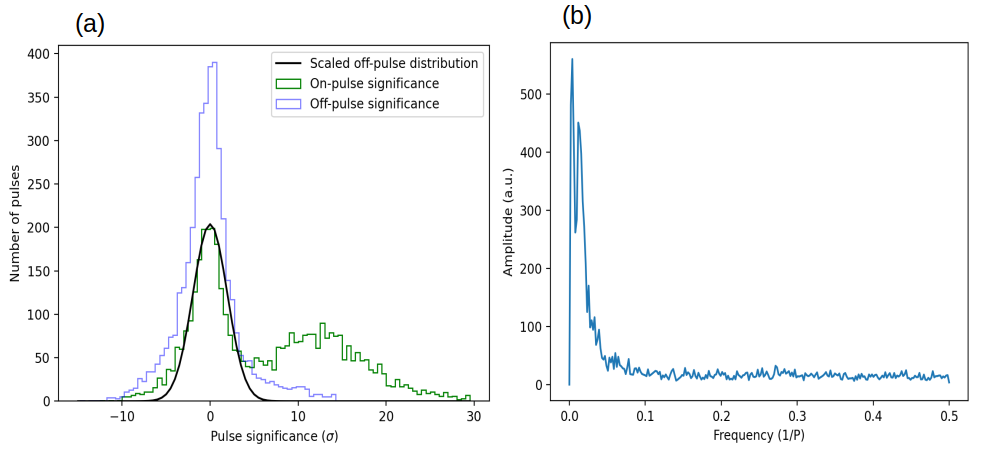}
    \caption{Histogram of the on and off pulse detection significance for the band-3 observation is shown in panel (a). Panel (b) shows the periodogram of the pulse sequence highlighting the quasi-periodicity of the nulling in this pulsar.}
    \label{fig:fig7}
\end{figure}

We use the dedispersed time series for the nulling analysis. First, we remove the baseline variations in the time series by subtracting a running median window matched according to the width of the profile. Then, we compute the on-pulse and off-pulse detection significance using an equal number of bins from the on and off-pulse phase for each pulse. We plot the histograms of on-pulse and off-pulse detection significance and use the method described by \citet{nulling_method} to calculate the nulling fraction of the pulsar. We fit a Gaussian shape in the off-pulse histogram to estimate its amplitude ($A_0$) and standard deviation ($\sigma$). We use this standard deviation to fit the negative part of the on-pulse energy histogram and estimate its amplitude ($A_1$) (see panel (a) of Fig. \ref{fig:fig7} and \ref{fig:fig9}). The nulling fraction is defined as $\frac{A_1}{A_0}$. The error in the nulling fraction is also calculated by using equation (3) of \citet{nulling_method}. We also stack the pulses with signal and pulses with nulls separately to check if there is any faint emission in the null region. We use a threshold value of 3$\sigma$ to classify between pulses with signal and pulses with nulls. We stack these classes separately and get a high significance profile of the pulsar in the cases of pulses with signal and a noise-like response in the case of nulls (see panel b of Fig. \ref{fig:fig6} and \ref{fig:fig8}). We also notice repeating patterns of nulling in the pulse sequences. To search and quantify any periodicity in the nulling, we prepare a sequence of ones and zeros as described by \citet{nulling_basu}. We put ones for pulses with signal and zeros for pulses with null in a sequence. Now, we subtract the mean of this sequence from itself to make it zero mean. Now, we take a block of 512 bins of the sequence and calculate the power spectra of this block by using the \texttt{periodogram} module from \texttt{scipy.signal} of python. Then we shift the block by 50 bins on the sequence and take the next block of 512 numbers and calculate its power spectra. We repeat this exercise until we reach the end of the sequence. We stack the power spectra obtained at each step to get the final periodogram.\\


\iffalse{\begin{figure}
   \centering
    \includegraphics[height=8 cm, width=12 cm]{2D_spectra.png}
    \caption{Time-varying power spectra of the zero-one series corresponding to the on and off pulses from the band-3 (300-500 MHz) data are shown in the upper panel. The plots have been zoomed around the frequency range of 0.0 to 0.1. This demonstrates the variations in the periodicity of the nulling with time. The lower panel shows the final power spectra after averaging in time.}
    \label{fig:2d_spectra}
\end{figure}}\fi
We used a band-3 uGMRT observation epoch with a duration of 1.6 hours to characterize the nulling in the 300 to 500 MHz frequency range. This observation has more than 4000 single pulses from this pulsar. Fig. \ref{fig:fig6} shows the pulse sequence (with an averaging of 8 pulses) with clear nulls from this observation epoch along with stacked profiles of pulses with signal and nulls. The stacked profile of nulls confirms that there is no low-level emission from the pulsar in the null phase. Fig. \ref{fig:fig7} shows the histogram of the on and off pulse significance along with the periodogram of the pulse sequence. The histogram of the off-pulse energy distribution shows long tails owing to the residual fast baseline variations. One can see clear bimodal distribution in the on-pulse energy distribution. The calculated nulling fraction is $57.1\pm 1.7 \%$. The final periodogram of the pulse sequence shows quasi-periodicity in the nulling. We see two time scales of quasi-periodicity in the periodogram, the first one peaks at around 250 pulses while the second time scale is around 64 pulses. The periodicity of nulling varies with time within the observation duration \citep{nulling_basu} and the width of the peaked structures in the final periodogram represents the variation in the periodicity. Though the two peaks in the final periodogram are not well separated with not enough frequency bin resolution, one can infer a rough estimate for the variations of the two timescales of periodicities considering the widths of peaked structures and the dip between the two peaks. The first periodicity timescale peaks at 250 periods and varies roughly between 130 to 500. The second periodicity timescale peaks at 70 periods and can vary between 40-100 periods. \\

We used a uGMRT band-4 observation to investigate the nulling in the frequency range 550$-$750 MHz. This observation is 1.3 hours long and contains more than 3300 pulses from the pulsar. Panel (a) of Fig. \ref{fig:fig8} shows the pulse sequence (with an averaging of 6 pulses) with clear nulls and panel (b) shows the profiles of pulses with signal and nulls from the pulsar and verifies that there is no faint emission in the null regions. The histogram of the on and off-pulse detection significance is shown in panel (a) of Fig. \ref{fig:fig9}. The nulling fraction calculated from this histogram is $55.1\pm 0.8 \%$. Panel (b) of the Fig. \ref{fig:fig9} shows the final periodogram of the pulse sequence with two time scales for the quasi-periodic nulling, similar to the quasi-periodicities we find in band-3 (as shown in Fig. \ref{fig:fig7}). This time, the first timescale peaks at the first bin of the periodogram, which corresponds to 512 pulses, and dips at the second bin which is 256 pulses. The second timescale however is consistent with the periodicities seen in band-3 analysis. It peaks at 64 periods and ranges between 50-130 periods.\\

We used seven short observations (each of $\sim$40 minutes) in band-3 (300-500 MHz) of this pulsar to robustly estimate the nulling fraction. These observations contain a total of 11883 pulses from this pulsar. Fig. \ref{fig:combined_null} shows the histogram of on and off-pulse energy of all these pulses. The nulling fraction estimated from this histogram is $63\pm 3.4\%$. The significant error in the estimate is due to a large error in fitting a Gaussian shape to the off-pulse energy distribution, as the distribution significantly deviates from the Gaussian shape due to the baseline variations in the time series of different epochs. Due to the short observation duration of individual epochs, we were not able to determine the quasi-periodicity of nulling in these observations.\\

\begin{figure}
   \centering
    \includegraphics[height=8 cm, width=18 cm]{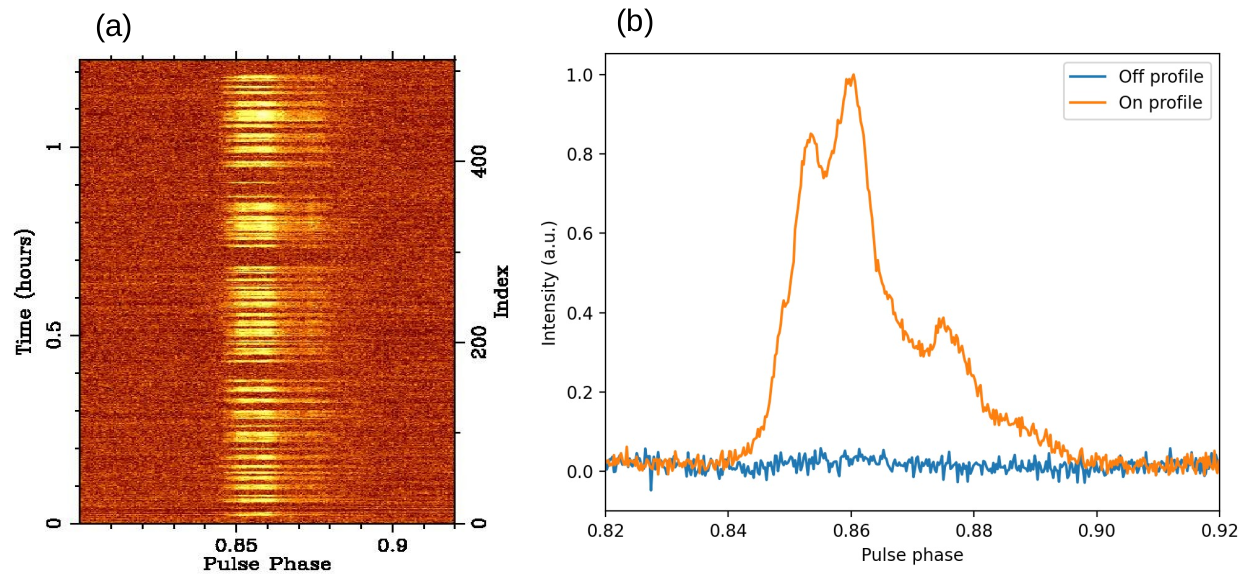}
    \caption{Nulling in band-4 (550$-$750 MHz) observation of uGMRT. Panel (a) shows the pulse sequence and the panel (b) shows the On and Off profile of this pulsar. The off profile obtained from the nulls does not have any significant signal.}
    \label{fig:fig8}
\end{figure}

\begin{figure}
   \centering
    \includegraphics[height=8 cm, width=18 cm]{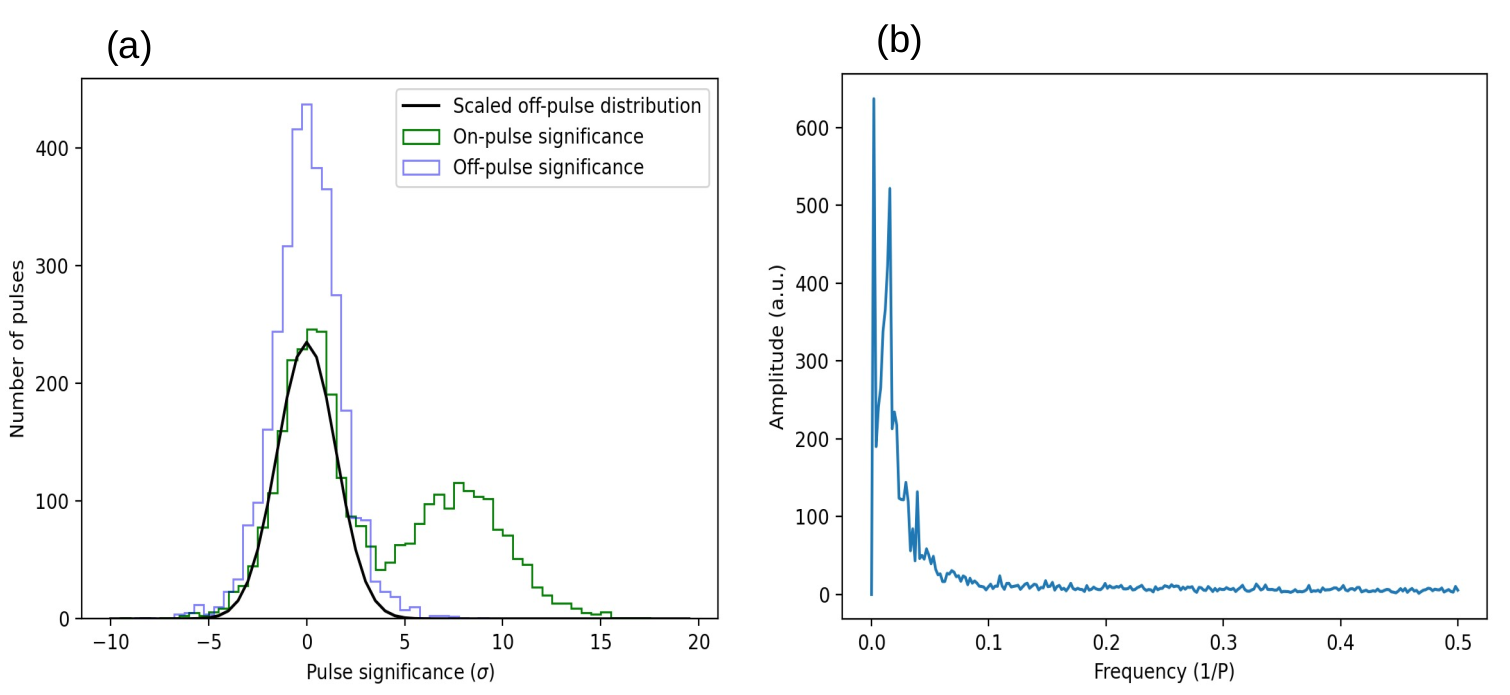}
    \caption{Panel (a) shows the histogram of the on and off pulse significance from the band-4 observation epoch. Panel (b) shows the periodogram of the pulse sequence.}
    \label{fig:fig9}
\end{figure}
\begin{figure}
   \centering
    \includegraphics[height=8 cm, width=14 cm]{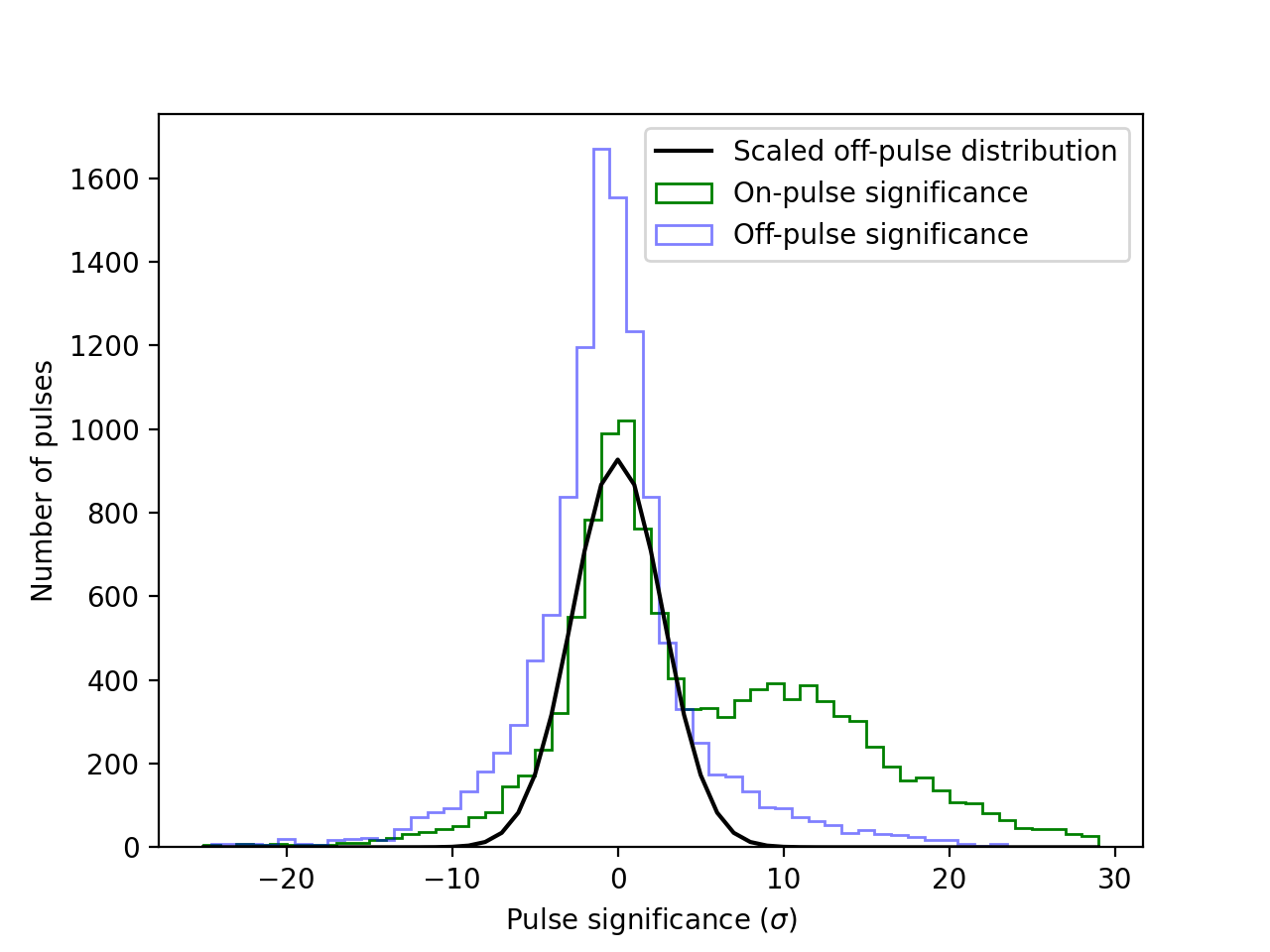}
    \caption{Histogram of on and off-pulse significance for 11883 pulses from seven epochs of observations with $\sim$40 minutes duration in band-3 (300$-$500 MHz). The estimated nulling fraction from this histogram is $63\pm3.4\%$.  }
    \label{fig:combined_null}
\end{figure}
\begin{figure}
   \centering
    \includegraphics[height=8 cm, width=14 cm]{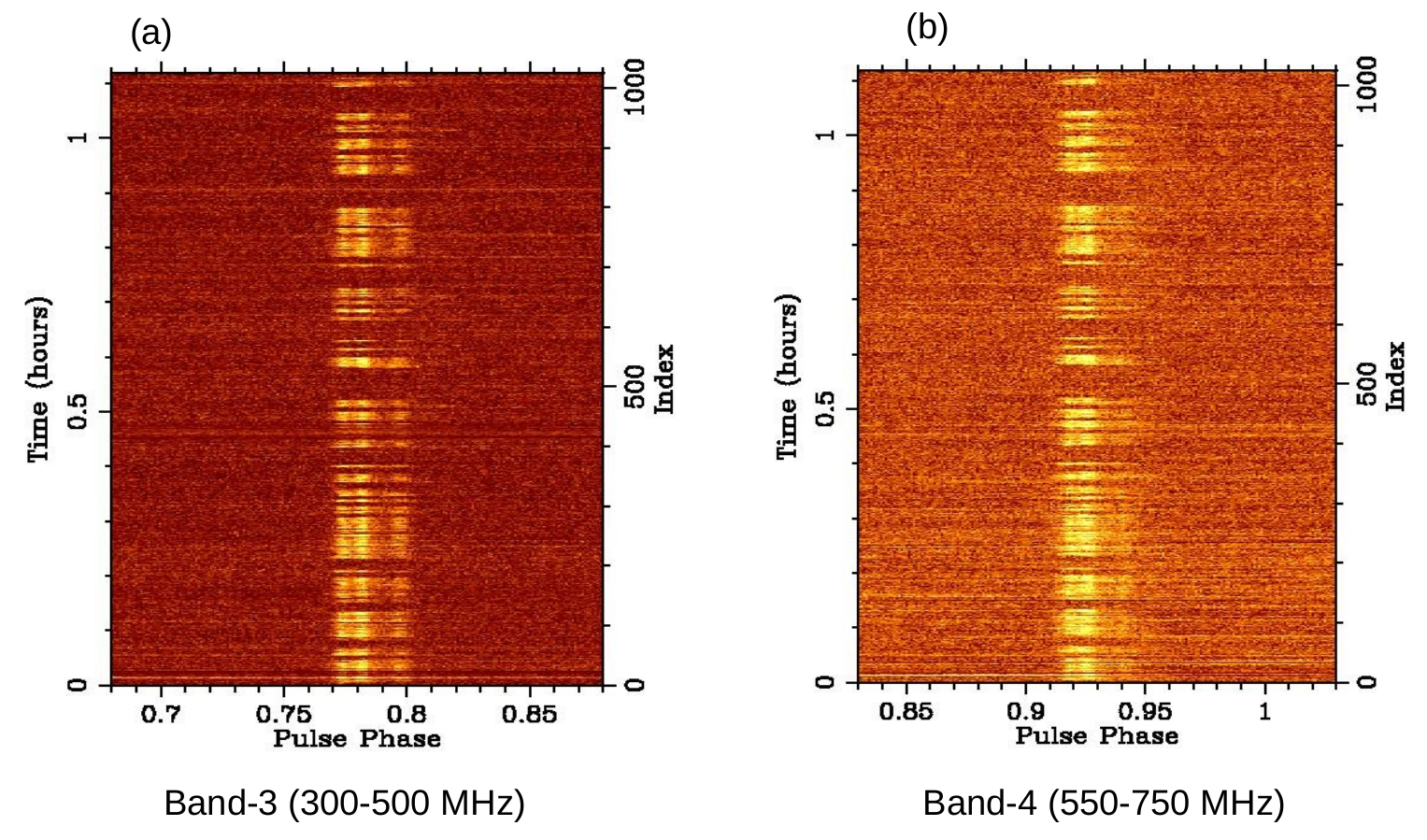}
    \caption{Pulse sequences from the simultaneous dual frequency observation. The pulses from band-3 (300$-$500 MHz) are shown in panel (a), while panel (b) shows the pulse sequence from band-4 (550$-$750 MHz) of uGMRT. These pulse sequences are visually identical.}
    \label{fig:fig10}
\end{figure}
\begin{figure}
   \centering
    \includegraphics[height=9 cm, width=15 cm]{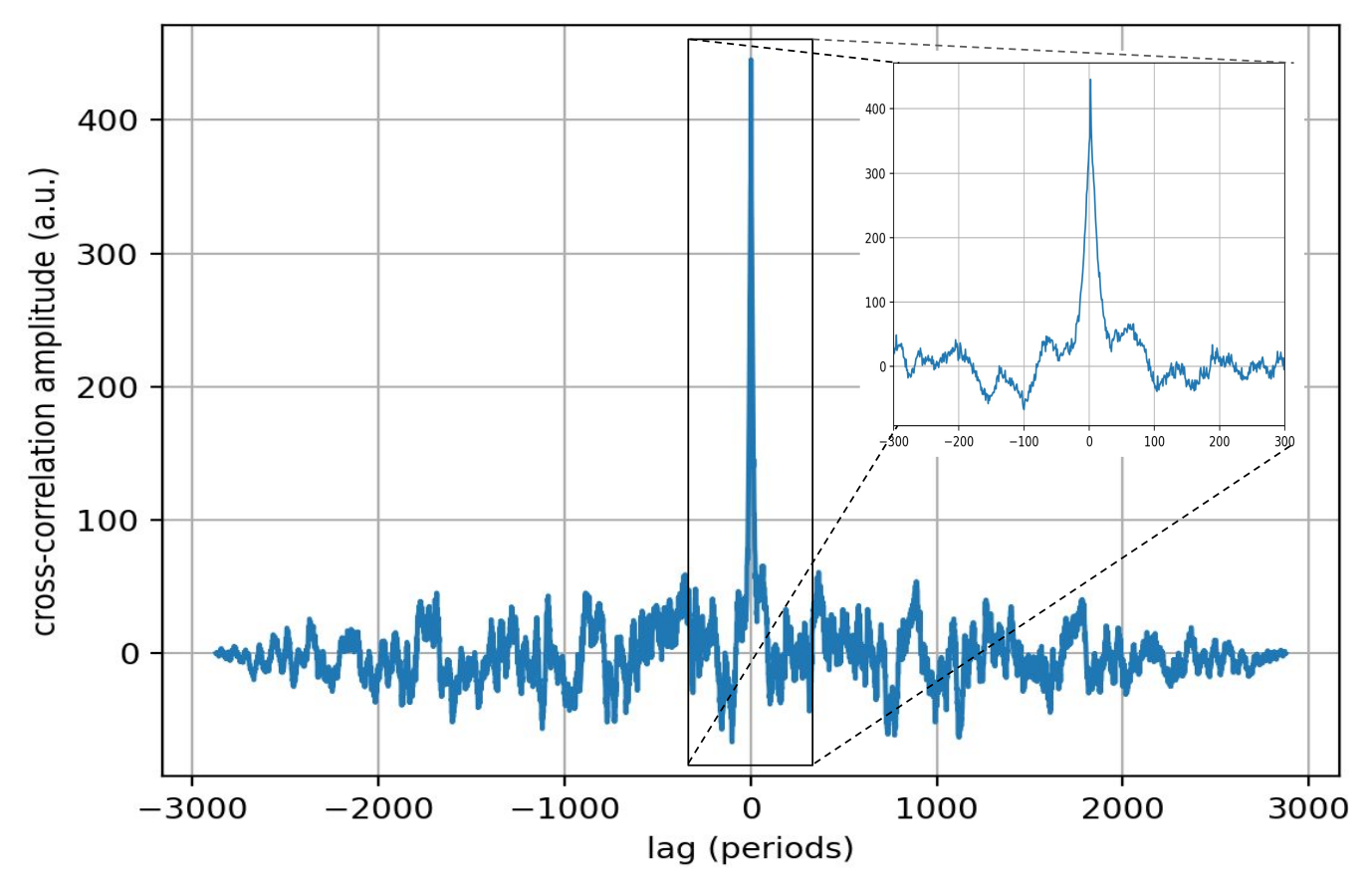}
    \caption{Cross-correlation function of the pulse sequences obtained from the two bands of the simultaneous dual frequency observation. A sharp peak at the zero lag is seen, verifying that these two pulse sequences are very similar. }
    \label{fig:fig11}
\end{figure}

We have estimated the nulling fractions and nulling periodicities in the two frequency bands, and they are slightly different. This may hint that nulling phenomenon is different in different frequencies, but a simultaneous observation in two frequency bands is required to truly examine the broadband nature of nulling. To demonstrate the broadband nature of the nulling in this pulsar, we use a simultaneous dual-frequency observation of this pulsar. The GMRT array was split into two subarrays. The subarray-1 having 10 antennas including 8 central square antennas and two arm antennas, was configured in band-3 (300$-$500 MHz), while the subarray-2 with 13 antennas, including four central square and nine arm antennas, was configured in band-4 (550$-$750 MHz). We observed the pulsar for 1.1 hours in this dual subarray mode. We got more than 2800 pulses from the pulsar simultaneously in both band-3 and band-4 of uGMRT. Now, we check for the simultaneous appearance of these pulses with signal and nulls in both the bands. Fig. \ref{fig:fig10} shows the pulse sequences (with an averaging of 3 pulses) from the two bands. The panel (a) shows pulses from band-3 while panel (b) represents the pulses from band-4. These two pulse sequences are very similar. The prominent bursts and nulls have roughly the same arrival time in both frequency bands. We generate the sequence of ones and zeros for these two pulse sequences, similar to that used in the periodicity analysis of nulling. We take the cross-correlation of these two sequences using the \texttt{correlate} module of \texttt{numpy} of python. Fig. \ref{fig:fig11} shows the cross-correlation function of these two sequences corresponding to two frequency bands. We see a sharp peak at zero lag, which is expected when the two pulse sequences are very similar. The zoomed version of the cross-correlation function also shows the width of the peak, which is close to $\sim 30$ pulses. This should correspond to the most common timescale of the burst in these observations. With this exercise, we successfully demonstrate the broadband nature of nulling in this pulsar.\\

To summarize, the pulsar J1244$-$4708 shows nulling with a nulling fraction of $\sim 55\%$ in both band-3 (300$-$500 MHz) and band-4 (550$-$750 MHz) of uGMRT. There are two timescales of quasi-periodicity of nulling in this pulsar, the first corresponding to a few hundred periods, while the second one corresponds to 40$-$130 periods and peaks at 60$-$70 periods. We also demonstrate the broadband nature of nulling using a simultaneous dual-frequency observation of this pulsar. We performed analysis for subpulse drifting using Longitude Resolved Fluctuating Spectra (LRFS)\citep{LRFS_basu} and did not find any other periodicity except the periodicity of nulling. In our observations, we also did not find a second mode of emission.

\section{Summary}\label{sec:sum}
Nulling pulsar emits irregular and sporadic signals with underlying periodicity. In some extreme cases like longer nulling pulsars and RRATs, The single pulse search seems to be more effective than the periodicity search. In this work, we show that any periodic signal can be better detected in a periodicity search than the single pulse search, irrespective of its nulling fraction, given that the length of the time series is sufficiently large. For example, GHRSS pointing of 10 mins duration can effectively find nulling pulsars with the nulling fraction equal to or less than $96\%$ from periodicity search, if the period of the pulsar is 1 second or smaller. This ensures that periodicity searches are useful in the search of periodic signals with very extreme nulling. A comparison of the FFA and FFT-based search methods over a range of nulling fractions shows that the FFA has an additional advantage over the FFT search for extremely nulling signals (having nulling fraction  $\sim$80\%). We also verify this trend by using data points from two nulling pulsars from the GHRSS survey (J1937$-$2937 and J1244$-$4708). So, the FFA search can be efficiently used to discover extreme nulling pulsars like the ones found in the GHRSS survey.\\

We report the localization and timing of a GHRSS pulsar J1244$-$4708 along with the frequency evolution of its profile and nulling properties. We find that the pulsar shows hints of inverse width-to-frequency relation, i.e. widths of the profile and its components are increasing with increasing frequency. The nulling fraction of this pulsar is close to $60\%$. The pulsar also shows quasi-periodicity in nulling with two timescales corresponding to $\sim 70$ periods and a few hundred periods respectively. \citet{nulling_basu} describe periodic nulling as an extreme form of amplitude modulation. They show that the pulsars with periodic nulling and amplitude modulation occupy a region on the modulation periodicity versus the spin-down energy ($\dot{E}$) plane that is separated from the region occupied by the pulsars showing subpulse drifting (see Fig. 3 of \citet{nulling_basu}). The measured values of nulling periodicities (70 periods and a few hundred periods) and the spin-down energy ($\sim 10^{31} ~erg~s^{-1}$) of the pulsar J1244$-$4708 puts it in the region occupied by pulsars exhibiting periodic nulling and amplitude modulation. We also demonstrate the broadband nature of nulling in this pulsar by using a simultaneous dual-frequency observation. We use the cross-correlation of the two pulse sequences to show that the pulse sequences from the two bands (band-3: 300$-$500 MHz and band-4: 550$-$750 MHz) of simultaneous dual frequency observation are very similar in order to verify the broadband nature of nulling. The fact that this pulsar shows quasi-periodicity in nulling and the nulling is seen simultaneously in both bands favours the cessation of coherent radio emission as the origin of nulling in this pulsar. This can arise either due to the cessation of pair production at the polar cap \citep{nulling_death_line} or loss of coherence due to temporarily unfavorable surface magnetic field configuration \citep{nulling_minhajur}. Given the period derivative of this pulsar, it being located in death valley seems unlikely and the loss of coherence in radio emission is likely to be a more favored mechanism for nulling in this pulsar.

\section{Acknowledgement}
We acknowledge the support of the Department of Atomic Energy, Government of India,  under project no.12-R\&D-TFR-5.02-0700.  The GMRT is run by the National Centre for Radio Astrophysics of the Tata Institute of Fundamental Research, India. We acknowledge the support of GMRT telescope operators and the GMRT staff for supporting the GHRSS survey observations.

\bibliography{sample631}{}
\bibliographystyle{aasjournal}

%% This command is needed to show the entire author+affiliation list when
%% the collaboration and author truncation commands are used.  It has to
%% go at the end of the manuscript.
%\allauthors

%% Include this line if you are using the \added, \replaced, \deleted
%% commands to see a summary list of all changes at the end of the article.
%\listofchanges

\end{document}